\documentclass[12pt,14paper]{article}
\usepackage{amsmath}
\usepackage{epsfig}
\usepackage{array}
\usepackage{float}
\usepackage{lscape,graphicx} 
\usepackage{graphics}
\usepackage{amssymb}
\usepackage{color}
\usepackage{multirow}
\usepackage{advdate}
\usepackage{datenumber}
\newcommand{\GeV}{\mbox{$~{\rm GeV}$}}

\newcommand{\bea}{\begin{equation}\begin{array}{c}}
\newcommand{\eea}{\end{array}\end{equation}}
\newcommand{\ea}{\end{array}}

\newcommand{\beq}{\begin{equation}}
\newcommand{\eeq}{\end{equation}}
\newcommand{\bad}{\begin{array}{ccc}}

\newcommand{\ba}{\begin{array}{c}}

\newcommand{\half}{\frac{1}{2}}

\newcommand{\sigmav}{\langle\sigma v\rangle}

\begin{document}

\title{
{\bf FIMP realization of the scotogenic model}}

\author{Emiliano Molinaro$^1\footnote{emiliano.molinaro@tum.de} $, Carlos E. Yaguna$^2\footnote{carlos.yaguna@uni-muenster.de}$, Oscar Zapata$^{1,3}\footnote{ozapata@fisica.udea.edu.co}$\\[7mm]
\it \normalsize $^1$Physik-Department T30d, Technische Universit\"at M\"unchen,\\ 
\it \normalsize James-Franck-Stra{\ss}e, 85748 Garching, Germany,\\[2mm]
\it \normalsize $^2$Institut f\"ur Theoretische Physik, Universit\"at M\"unster,\\
\it \normalsize Wilhelm-Klemm-Stra\ss e 9, D-48149 M\"unster, Germany,\\[2mm]
\it \normalsize  $^3$Instituto de F\'{i}sica, Universidad de Antioquia,\\
\it \normalsize  A.A. 1226, Medell\'{i}n, Colombia
}
\date{}
\maketitle
\vspace*{-12cm}
\begin{flushright}
\texttt{\footnotesize FLAVOUR(267104)-ERC-67}\\[-1mm] 
\texttt{\footnotesize TUM-HEP 941/14}\\[-1mm] 
\texttt{\footnotesize NORDITA-2014-55}\\[-1mm] 
\texttt{\footnotesize MS-TP-14-19}
\end{flushright}
\vspace*{8.5cm}
\thispagestyle{empty}
\begin{abstract}
The scotogenic model is one of the simplest scenarios for physics beyond the Standard Model that can  account for neutrino masses and dark matter at the TeV scale. It contains another scalar doublet and three additional singlet fermions ($N_i$), all odd under a $Z_2$ symmetry. In this paper, we examine the possibility that the dark matter candidate, $N_1$, does not reach thermal equilibrium in the early Universe so that it behaves  as a Feebly Interacting Massive Particle (FIMP). In that case, it is found that the freeze-in production of dark matter  is entirely dominated by the decays of the odd scalars. We compute the resulting dark matter abundance and study its dependence with the parameters of the model. The freeze-in mechanism is shown to be able to account for the observed relic density over a wide range of dark matter masses, from  the keV to the TeV scale. In addition to freeze-in, the $N_1$ relic density  receives a further contribution from the late decay  of the next-to-lightest odd particle, which we also analyze. Finally, we  consider the possibility that  the dark matter particle is a WIMP but receives an extra contribution to its relic density from the decay of the FIMP ($N_1$). In this case, important signals at direct and indirect detection experiments are generally expected.  
\end{abstract}

\section{Introduction}
The identification of the dark matter particle stands as  one of the most pressing problems in fundamental physics today. Its solution  requires physics beyond the Standard Model but it is not yet known what this new physics is. Most of the models studied in the literature assume that the dark matter consists of  Weakly Interacting Massive Particles (WIMPs) and that its relic density is the result of a freeze-out process.  One  advantage of this scenario is that it naturally yields a relic density of the same order as the observed dark matter density --the so-called WIMP-miracle.  In addition, WIMP models generally give rise to  signals, in direct and indirect detection experiments as well as at colliders such as the LHC, that are within the reach of current experiments. Up to now, however, such signals have not been found and strong bounds on many of these models have been derived.  If this situation persists for the next few years, the WIMP paradigm would likely have to be abandoned \cite{Bertone:2010at} and dark matter would have to be explained in some other way.  

A simple and  appealing alternative to the WIMP framework is provided by FIMP (Feebly Interacting Massive Particle) dark matter \cite{Hall:2009bx}. Its basic idea is that, in contrast to WIMPs,  the dark matter interacts so weakly that it does not reach thermal equilibrium in the early Universe. Thus, its relic density is not the result of a freeze-out. Instead, the dark matter particles are  slowly produced via decays or scatterings of the particles in the thermal plasma --a process dubbed \emph{freeze-in}-- but they are never abundant enough for their annihilations to be relevant. Consequently, the dark matter abundance steadily increases as the Universe cools down  until the so-called freeze-in temperature is reached, and it remains constant afterward. Due to its feeble interactions, FIMPs do not give rise to observable signals neither in direct nor in indirect dark matter detection experiments. These experiments, therefore,  provide an unambiguous way of testing or falsifying this scenario: if a signal were detected, one could immediately conclude that dark matter does not consist of FIMPs. Currently, FIMPs provide a viable and attractive framework to account for the dark matter.

Several explicit realizations of the FIMP framework have already been investigated \cite{McDonald:2001vt,Asaka:2005cn,Asaka:2006fs,Kusenko:2006rh,Cheung:2011nn,Yaguna:2011qn,Yaguna:2011ei,Blennow:2013jba,Chu:2013jja,Merle:2013wta,Klasen:2013ypa}. FIMPs are necessarily singlets under the Standard Model (SM) gauge group so the two simplest extensions of the SM that incorporate  a FIMP include a new singlet scalar \cite{McDonald:2001vt,Yaguna:2011qn,Yaguna:2011ei} or  a new singlet fermion \cite{Klasen:2013ypa}, both of which give rise to an interesting phenomenology. In this paper, we will study a richer realization of the FIMP framework based on the scotogenic model (also known as the radiative seesaw model) \cite{Ma:2006km}.  This model is one of the simplest scenarios for physics beyond the SM that can simultaneously account for neutrino masses and dark matter at the TeV scale. It contains another scalar doublet and three additional singlet fermions ($N_i$), all odd under a $Z_2$ symmetry.  Even though the phenomenology of this model has been extensively studied in a number of previous works --see e.g.  \cite{Kubo:2006yx,Sierra:2008wj,Gelmini:2009xd,Suematsu:2009ww,Suematsu:2011va,Schmidt:2012yg,Hu:2012az,Kashiwase:2012xd,Kashiwase:2013uy}--,  none of them considered the possibility of FIMP dark matter. The basic idea is that the couplings of one of the singlet fermions, $N_1$, are so small  that it does not reach thermal equilibrium in the early Universe and is instead produced via freeze-in. We show that dark matter production is dominated by  the decays of the odd scalars and  study the dependence of the resulting abundance with the different parameters of the model. In particular, the viable parameter space for FIMP dark matter is precisely determined and it is shown to span a wide range of dark matter masses, from the keV to the TeV scale. Besides freeze-in, the dark matter relic density receives an additional contribution from the so-called  superWIMP mechanism \cite{Feng:2003xh} which strongly depends on the identity of the  next-to-lightest odd particle.  We identify an important region of the parameter space where this contribution is always negligible and freeze-in production is dominant.

Another interesting setup we discuss occurs when the dark matter particle (the lightest odd particle) is not $N_1$ but $H^0$, a WIMP.  In that case, $N_1$, which is produced via freeze-in,  decays into the dark matter, increasing its relic density and allowing for new viable regions in the parameter space.  We show that interesting signals  from direct and indirect detection experiments are generally expected in this configuration.

The rest of the paper is organized as follows. In the next section, we introduce the model and discuss the experimental bounds it is subject to. Then in section \ref{sec:out} we obtain the conditions
necessary to ensure that $N_1$ does not reach thermal equilibrium in the early Universe. Our main results are presented in sections \ref{sec:fimp} and \ref{sec:decay}. In the former, we carefully study the production of FIMP dark matter and obtain the corresponding viable parameter space.  Section \ref{sec:decay} is dedicated to the case where the dark matter particle is $H^0$ and receives a contribution to its relic density from FIMP decays. Finally, we present our conclusions in section \ref{sec:con}.

\section{The model}

The model we consider is the so-called scotogenic model \cite{Ma:2006km}, one of the simplest models that can simultaneously explain neutrino masses and dark matter at the TeV scale. In it, the SM is extended with  a second Higgs doublet $H_{2}\equiv(H^{+},\,H_2^0)$
and three Majorana neutrinos $N_{j}$ ($j=1,2,3$), all odd under an exact $Z_{2}$ symmetry (the SM fields are instead even under it). This symmetry forbids the coupling between $H_2$ and the quark fields, which would give rise to flavor changing neutral currents, and it guarantees the stability of the dark matter particle.

The Lagrangian of the model contains the following new terms involving the singlet fields
\begin{eqnarray}
\mathcal{L} &\supset &
Y^{\nu}_{\alpha i}\,\left(\overline{\nu}_{\alpha L}\,H_{2}^{0}\,-\,\overline{\ell}_{\alpha L}\,H^{+}\right)\,N_{i}
	 +\,\half\,M_{j}\,\overline{N}_{j} \,N_{j}^{C}\,+\,{\rm H.c.}\label{modlagr}
\end{eqnarray}
Hence, the singlets have Majorana masses $M_j$ and interact only with $H_2$ and the lepton doublets. The most general scalar potential of this model is given by
\begin{eqnarray}
	V(H_{1},H_{2}) &=& -\,\mu_{1}^{2}\,\left(H_{1}^{\dagger}\,H_{1}\right)\,+\,\lambda_{1}\,\left(H_{1}^{\dagger}\,H_{1}\right)^{2}\,+\,\mu_{2}^{2}\,\left(H_{2}^{\dagger}\,H_{2}\right)\,+\,\lambda_{2}\,\left(H_{2}^{\dagger}\,H_{2}\right)^{2} \nonumber\\
	&&\,+\,\lambda_{3}\,\left(H_{1}^{\dagger}\,H_{1}\right)\,\left(H_{2}^{\dagger}\,H_{2}\right)
	\,+\,\lambda_{4}\,\left(H_{1}^{\dagger}\,H_{2}\right)\,\left(H_{2}^{\dagger}\,H_{1}\right)\nonumber\\
	&&\,+\, \frac{\lambda_{5}}{2}\,\left[\left(H_{1}^{\dagger}\,H_{2}\right)^{2}\,+\,{\rm H.c.}\right]\,,
\end{eqnarray}
where $\mu_{1,2}^2>0$ and $H_1$ is the SM Higgs doublet. It is convenient to write $H_2^{0}=(H^0+iA^0)/\sqrt 2$ as the $\lambda_5$ term in the Lagrangian creates a mass-splitting between $H^0$ and $A^0$. After electroweak symmetry breaking, $\langle H_{1}\rangle =(0, v/\sqrt{2})$ with  $v\simeq 246$ GeV, the scalar spectrum consists of one $Z_2$ even field ($H$, the SM Higgs boson recently discovered at the LHC with a mass of $125$ GeV \cite{Aad:2012tfa,Chatrchyan:2012ufa}) and four $Z_2$ odd particles:
\begin{itemize}
	\item A CP-even neutral scalar $H^{0}$  with mass $m_{H^{0}}^{2}=\mu_{2}^{2}
			\,+\,v^{2}\,\left(\lambda_{3}+\lambda_{4}+\lambda_{5}\right)/2$.
	\item A CP-odd neutral scalar $A^{0}$ with mass $m_{A^{0}}^{2}=\mu_{2}^{2}
			\,+\,v^{2}\,\left(\lambda_{3}+\lambda_{4}-\lambda_{5}\right)/2$.
	\item Two charged scalars $H^{\pm}$ with masses $m_{H^{\pm}}^{2}=\mu_{2}^{2}\,+\,v^{2}\,\lambda_{3}/2$.	
\end{itemize}
The free parameters of the model can be taken to be the masses of all the odd particles ($M_k$, $m_{H^0}$,$m_{A^{0}}$, $m_{H^{\pm}}$), two quartic couplings $\lambda_{2}, \lambda_L\equiv (\lambda_{3}+\lambda_{4}+\lambda_{5})/2$, and the set of $9$ Yukawa couplings ($Y^{\nu}_{\alpha i}$), which for simplicity we take to be real.  As explained below, these parameters are subject to a number of phenomenological constraints. In our numerical estimates, we will often use $y_k$ ($k=1,2,3$) to denote a typical value for the Yukawa coupling associated with the singlet $N_k$, $Y^{\nu}_{ik}\sim y_k$. We will also assume, following the spirit of this model, that the masses of the odd particles all lie at or below the TeV scale.

Notice that this model includes the well-known inert doublet model \cite{LopezHonorez:2006gr,Barbieri:2006dq,Honorez:2010re,LopezHonorez:2010tb,Goudelis:2013uca,Garcia-Cely:2013zga,Arhrib:2013ela} but has a more interesting phenomenology. It can explain neutrino masses, it gives rise to lepton-flavor violating processes \cite{Kubo:2006yx, Toma:2013zsa}, it contains  another dark matter candidate, it can realize thermal leptogenesis \cite{Suematsu:2011va,Kashiwase:2012xd,Hambye:2009pw,Racker:2013lua}, and it allows for  new effects on the relic density \cite{Klasen:2013jpa}. From the inert doublet model, it inherits several features, including the bounds on the masses of the odd scalar particles. They read $m_{H^0}+m_{A^0}>M_Z$ from the $Z$-width measurement, and $\mathrm{max}[m_{H^0},m_{A^0}]\gtrsim 100~\mathrm{GeV}$ \cite{Lundstrom:2008ai} and $m_{H^+}>70\mathrm{-}90~\mathrm{GeV}$ \cite{Pierce:2007ut} from collider searches at LEP.

In this model, neutrinos acquire Majorana masses  via 1-loop diagrams mediated by the odd particles. The resulting   light neutrino mass matrix is given by
\begin{align}
	\left(\mathcal{M}_{\nu}\right)_{\alpha \beta} &~= \sum_{k}\,\frac{Y^{\nu}_{\alpha k}\,Y^{\nu}_{\beta k}}{16\,\pi^{2}}\,M_{k}\,
	\left[\frac{m_{H^{0}}^{2}}{m_{H^{0}}^{2}-M_{k}^{2}}\log\left(\frac{m_{H^{0}}^{2}}{M_{k}^{2}}\right)\,-\,
	\frac{m_{A^{0}}^{2}}{m_{A^{0}}^{2}-M_{k}^{2}}\log\left(\frac{m_{A^{0}}^{2}}{M_{k}^{2}}\right)\right]\nonumber\\
	&\overset{^{\lambda_{5}\ll 1}}{=} \frac{\lambda_{5}\,v^{2}}{16\,\pi^{2}}\,\sum_{k}\,Y^{\nu}_{\alpha k}\,Y^{\nu}_{\beta k}\,\frac{M_{k}}{m_{0}^{2}-M_{k}^{2}}\,
	\left(1-\frac{M_{k}^{2}}{m_{0}^{2}-M_{k}^{2}}\,\log\left(\frac{m_{0}^{2}}{M_{k}^{2}}\right)\right)\,,
\label{eq:mn}
\end{align}
where  $m_{0}^{2}=\left(m_{H^{0}}^{2}+m_{A^{0}}^{2}\right)/2$ and we used $m_{H^{0}}^{2}-m_{A^{0}}^{2}=\lambda_{5}\, v^{2}$. As we will see in the next section, the out of equilibrium condition forces the Yukawa couplings of $N_1$, $Y^\nu_{i1}$, to be  so small that they give a negligible  contribution to neutrino masses. Effectively, then, $N_1$ decouples from neutrino masses and the sum in equation~(\ref{eq:mn}) is only over $k=2,3$. In consequence, only two light neutrinos acquire non-zero masses in this setup. This result is not generic to the scotogenic model but follows instead from the  requirement we have imposed of preventing one particle from reaching thermal equilibrium in the early Universe. 

Let us now estimate analytically the range of couplings that gives rise to viable neutrino masses.  
If $m_{0}^2\ll M_{k}^2$ we can simplify equation (\ref{eq:mn}) and write
\begin{align}\label{eq:numass}
\left(\mathcal{M}_{\nu}\right)_{\alpha \beta} =&\frac{\lambda_5\,v^2}{16\,\pi^2}\sum_k\frac{Y^\nu_{\alpha k}Y^\nu_{\beta k}}{M_k}\left(\ln\frac{M_k^2}{m_0^2}-1\right)\\
\approx & 10^{-2}\mbox{eV}\left(\frac{\lambda_5\, y_{2,3}^2}{10^{-11}}\right)\left(\frac{1\,\mbox{TeV}}{M_{2,3}}\right).
\label{eq:numass2}
\end{align}
Whereas for $m_{0}^2\gg M_{k}^2$ we get instead
\begin{align}\label{eq:numassB}
\left(\mathcal{M}_{\nu}\right)_{\alpha \beta} =&\frac{\lambda_5\,v^2}{16\,\pi^2m_0^2}\sum_k Y^\nu_{\alpha k}Y^\nu_{\beta k} M_k\\
\approx & 10^{-2}\mbox{eV}\left(\frac{\lambda_5\, y_{2,3}^2}{10^{-11}}\right)\left(\frac{1\,\mbox{TeV}}{m_0}\right)\left(\frac{M_{2,3}}{m_0}\right).
\label{eq:numass2B}
\end{align}

The above expressions tell us that, if we want to generate light neutrino masses with new physics at the TeV scale, the product $\lambda_5\, y_{2,3}^2$ must necessarily be very small ($\sim10^{-11}$). This condition can be satisfied in different ways, however. One can, for instance, set $\lambda_5\sim 10^{-9}$ so that $y_{2,3}\sim 0.1$. Or one could fulfill it with $\lambda_5=10^{-1}$ and $y_{2,3}\sim 10^{-5}$.  Moreover, we can  use equation~(\ref{eq:numass2}) and (\ref{eq:numass2B}) to set a lower bound on the Yukawa couplings:
\beq
y_{2,3}\gtrsim 10^{-6}.
\label{eq:lower}
\eeq
Smaller values of  $y_{2,3}$ would fail to reproduce the observed neutrino mass scale.

Since  equation~(\ref{eq:mn}) has the same matrix structure as the usual seesaw equation, one can adapt the Casas-Ibarra parametrization \cite{Casas:2001sr} to it and express the Yukawa couplings in terms of the experimental data on neutrino masses and mixing angles. For the analogous case of two-right handed neutrinos which is relevant in our scenario, this procedure introduces only one free parameter \cite{Ibarra:2003up,Ibarra:2011xn}, an angle that we take to be real. We assume a normal hierarchical spectrum  for the neutrinos and took their oscillation parameters from \cite{Capozzi:2013csa}. In this way, we guarantee that all the models we  consider in the following are compatible with current neutrino data.

The same interactions that generate neutrino masses induce lepton flavor violating processes such as $\mu\to e\,\gamma$ and $\tau\to \mu\,\gamma$ at the 1-loop level \cite{Kubo:2006yx,Toma:2013zsa}. Since these processes have not been observed, one must ensure that the predicted branching ratios are below the present experimental bounds. Given that the current limits read $\text{BR}(\mu\to e\,\gamma)<5.7\times 10^{-13}$ \cite{Adam:2013mnn} and $\text{BR}(\tau\to \mu\,\gamma)<4.4\times 10^{-8}$ \cite{Aubert:2009ag}, the former decay typically gives a stronger bound.
 In this model, the branching ratio for the $\mu\to e\gamma$ process is \cite{Kubo:2006yx}
\begin{eqnarray}
	\text{BR}(\mu\to e\,\gamma) & = & \frac{3\alpha_{\rm em}}{64\,\pi\,\left(G_{F}\, m_{H{^\pm}}^{2}\right)^{2}}\,\left| Y^{\nu}_{\mu k}\, Y{^{\nu}_{e k}}^{*}\, F_{2}\left(\frac{M_{k}^{2}}{m_{H^{\pm}}^{2}}\right) \right|^{2}\,\label{eq:brmueg}\\
	&\approx& 10^{-15}\left(\frac{100\,\mbox{GeV}}{m_H^{\pm}}\right)^4\,\left|\frac{y_{2,3}}{10^{-2}}\right|^4\,\left(\frac{F_2(M_{2,3}^2/m_{H^{\pm}}^2)}{3\times10^{-3}}\right)^2\,,\label{meg1}
\end{eqnarray}
where the loop function $F_2(x)$ varies in the range $[3\times 10^{-5},\,0.14]$ for $x=[10^{4},\,0.1]$  
and we have already taken into account the fact  that $y_1$ is negligible. Notice, in particular, that large Yukawa couplings,  $y_{2,3}\gtrsim 0.1$, are strongly disfavored. 
In our analysis, we always impose that $\text{BR}(\mu\to e\gamma)$, computed from equation~(\ref{eq:brmueg}), be below the experimental limit.

Another important bound that must be taken into account is the dark matter constraint --the requirement that the predicted relic density agrees with the observed dark matter density. In this model there are two viable dark matter candidates, the lightest neutral scalar and the lightest singlet fermion, and the predicted relic density depends on how they were produced in the early Universe. While most previous works have assumed the usual freeze-out scenario,  we want to examine the possibility that $N_1$ does not reach thermal equilibrium in the early Universe and is instead produced via freeze-in. 

\section{Out of equilibrium conditions}
\label{sec:out}

The basic requirement of the FIMP (or freeze-in) mechanism is that the dark matter particle does not reach thermal equilibrium in the early Universe. In the scotogenic model, only the fermions, which are gauge singlets, can play the role of FIMPs. Equilibrium can be prevented if their Yukawa interactions are sufficiently suppressed. In this section we analyze the different processes that can produce singlets and obtain the conditions necessary for them not to reach equilibrium. In particular we show that only one of them, denoted by $N_1$, can play the  role of a FIMP. 

Because the singlet fermions have Yukawa interactions of the form $N_{k}\,L \,H_2$, they can be produced via the two-body decay of the odd scalars.  The decay rate for the production of $N_{1}$ is approximately given by $\Gamma(H_2\to N_1\,L)=M_{H_2}\,y_1^2/(8\pi)$ --see equations~(\ref{GS1}) and (\ref{GS2}) below. Then, the out of equilibrium condition for this decay reads
\beq
\Gamma(H_2\to N_1\,L)\lesssim H(T\sim M_{H_2})\,,
\eeq
which for $M_{H_2}\sim100\GeV$ implies
\beq
y_1\lesssim 10^{-8}.
\label{eq:out}
\eeq
If $y_1$ were larger than this value, $N_1$ would be produced abundantly enough to reach thermal equilibrium. This small value of the Yukawa coupling implies that, as already anticipated in the previous section,  $N_1$ gives a negligible contribution to neutrino masses --see equation (\ref{eq:lower}). The heavier singlets, $N_{2,3}$, can be produced either via scalar decays ($H_2\to N_{2,3}\,L$) or, if they are heavier than the scalars, via the inverse decay  $H_2+L\to N_{2,3}$, both of which are in equilibrium for $y_{2,3}\gtrsim 10^{-8}$. Since the bound from neutrino masses requires $y_{2,3}\gtrsim 10^{-6}$, we can conclude that   $N_2$ and $N_3$ necessarily reach thermal equilibrium in the early Universe.  This model, therefore, admits only one FIMP: $N_1$. 

$N_1$ can also be produced via the decay of the heavier singlets or via $2\to 2$ scatterings of SM leptons or odd scalars. All these processes are however subdominant and do not modify the equilibrium condition obtained above. 

Contrary to our findings, it was stated in \cite{Gelmini:2009xd} that all three singlets could be out of equilibrium while explaining neutrino masses. The reason for this erroneous conclusion is that they failed to recognize the importance of the scalar decays as a production process for the singlets. Instead, they assumed that singlets were pair-produced via the annihilation of two leptons or two odd scalars. Since the rates of these processes depend on the neutrino Yukawa couplings to the fourth power (rather than the second), the out of equilibrium condition gives the wrong (and weaker) bound 
$y_i\lesssim10^{-4}$, which is consistent with the limit from neutrino masses.  From our discussion, it should be clear  though    that the decays of the odd scalars (or the inverse decays mentioned above) cannot be neglected as they are the dominant production process for the singlets. Once these decays are taken into account it follows that only one singlet can be out of equilibrium.    

Summarizing, the bound from neutrino masses implies that in this minimal setup (with three singlet fermions) only one FIMP is allowed. We will next show that this FIMP can easily account for the observed dark matter density via freeze-in. 

\section{FIMP dark matter}
\label{sec:fimp}
In this section we analyze the case where the dark matter candidate --that is the lightest odd particle-- is the singlet that does not reach thermal equilibrium in the early Universe, which we denote by $N_1$. That is, we consider the spectrum $M_1<M_{2,3},\,m_{H^0},\,m_{A^0},\,m_{H^\pm}$. The $N_1$ relic density, $\Omega_{N_{1}}\,h^{2}$, will therefore have two contributions: one from the freeze-in mechanism and another one from the late decay of the next-to-lightest odd particle, which we call the superWIMP contribution \cite{Feng:2003xh}. We have therefore
\beq
\Omega_{\text{DM}}\,h^{2}\equiv\Omega_{N_{1}}\,h^{2}=\Omega^{freeze-in}\,h^{2}+\Omega^{superWIMP}\,h^{2}.
\eeq 
Since these two contributions are entirely independent --they become relevant at different temperatures and do not depend on the same  parameters-- we will study them separately. 

\subsection{The freeze-in contribution}
Let us discuss first the freeze-in contribution to $N_1$ production. Since $N_1$ has a direct coupling to leptons and to odd scalars, its production will be dominated by the decays of the scalars ($H^0,A^0,H^\pm$) while they are   in equilibrium with the thermal bath. The $N_1$ yield, $Y_{N_{1}}(T)=n_{N_1}(T)/s(T)$, is computed by solving 
the following Boltzmann equation \cite{Hall:2009bx}
\begin{equation}
	s\, T\,\frac{d Y_{N_{1}}}{dT} \; = \; -\frac{\gamma_{N_{1}}(T)}{H(T)}\,,\label{BE}
\end{equation}
where $s$ is the entropy density of the Universe, $H(T)$ is the expansion rate of the Universe at a given temperature and $\gamma_{N_{1}}(T)$ is the thermal averaged FIMP production rate. We have that 
\begin{equation}
	\gamma_{N_{1}}(T)\;=\; \sum\limits_{X}\frac{g_{X}\,m_{X}^{2}\, T}{2\,\pi^{2}}\,K_{1}\left(m_{X}/T\right)\,\Gamma\left(X\to N_{1} \, \ell\right),
\end{equation}
where $X=H^0,A^0,H^\pm$  and $\ell$ is a SM lepton. In this equation, $K_{1}(x)$ is the Bessel function of the second kind, and $g_{X}$ is the number of internal degrees of freedom of particle $X$.
Specifically, $g_{H^{0},A^{0},H^+,H^-}=1$. The decay rates that enter into this expression are calculated as 
\begin{eqnarray}
\label{GS1}
	\Gamma\left(H^{0}/A^{0}	\to N_{1} \, \nu_{\alpha}  \right) & = & \frac{m_{H^{0}/A^{0}}\,\left| Y^{\nu}_{\alpha 1} \right|^{2}}{32\,\pi}\left(1-\frac{M_{1}^{2}}{m_{H^{0}/A^{0}}^{2}}\right)^{2}\,\approx \frac{m_{H^{0}/A^{0}}\,\left| Y^{\nu}_{\alpha 1} \right|^{2}}{32\,\pi},\\
	\Gamma\left(H^{+}	\to N_{1} \, \overline{\ell_{\alpha}}  \right) & = & \frac{m_{H^{+}}\,\left| Y^{\nu}_{\alpha 1} \right|^{2}}{16\,\pi}\left(1-\frac{M_{1}^{2}}{m_{H^{+}}^{2}}\right)^{2}\,\approx \frac{m_{H^{+}}\,\left| Y^{\nu}_{\alpha 1} \right|^{2}}{16\,\pi}.\label{GS2}
\end{eqnarray}
where the approximations are valid unless there is a strong mass degeneracy between $N_1$ and one of the scalars. 

It is easy to verify that the decays of the heavier singlet fermions, $N_{2,3}\to N_1\,\bar{\ell}\,\ell$, give a negligible contribution to dark matter production. In fact, the corresponding decay rate is given by 
\begin{equation}
	\Gamma(N_{2,3} \to N_{1}  \,\overline{\ell_{\alpha}}\, \ell_{\beta}) 
	\,=\, \frac{M_{2,3}^{5}}{6144\,\pi^{3}\,m_{S}^{4}}\left(\left|Y^{\nu}_{\beta 1} \right|^{2} \left|Y^{\nu}_{\alpha 2,3} \right|^{2}+\left|Y^{\nu}_{\alpha 1} \right|^{2} \left|Y^{\nu}_{\beta 2,3} \right|^{2}\right)\,,
\end{equation}
which is always much smaller than (\ref{GS1}) and (\ref{GS2}).  Other negligible  processes are the production of dark matter  via scatterings of two $Z_{2}$-odd particles or two SM particles. Both are  always subdominant because the corresponding cross-sections are proportional to the fourth power of the  Yukawa couplings. Thus, the $N_1$ abundance, $Y_{N_{1}}$, is solely determined by the Yukawa couplings ($Y^{\nu}_{\alpha 1}$) and by the spectrum of  odd scalar particles: $m_{H^0}$, $m_{A^0}$, $m_{H^\pm}$. As we will see, for our purposes it is often a good approximation to consider all odd scalars to be degenerate, in which case we denote their common mass by $m_S$.  

From equations~(\ref{BE}--\ref{GS2}), taking into account that $s(T)=2\pi^2g_s T^{3}/45$, $H(T)=1.66\sqrt{g_\rho} T^{2}/M_{Pl}$ and $K_{1}(x)\sim1/x$ for $x\ll 1$, we have at high temperatures $T> m_{S}$:
\begin{equation}
	\frac{d Y_{N_{1}}}{dT} \; \approx \; - \,5\times 10^{3}\,\text{GeV}^{3}\,\left(\frac{m_{S}}{1\,\text{TeV}}\right)^{2}\,\left(\frac{y_{1}}{10^{-8}}\right)^{2}\,T^{-4}\,.\label{dYapprox}
\end{equation}
Therefore, on the one hand we have that at $T> m_{S}$ the yield always scales as the square of the scalar masses and of the $N_1$ Yukawa couplings. On the other hand,
at $T\lesssim m_{S}$ the scalar particle abundance becomes Boltzmann suppressed  and  the production of  dark matter is no longer efficient. As a result we have
\begin{equation}
	Y_{N_{1}}\left(T\lesssim m_{S} \right)\;\approx\;  10^{-4}\,\left(\frac{1~\text{TeV}}{m_{S}}\right)\,\left(\frac{y_{1}}{10^{-8}}\right)^{2}\,.\label{Yapprox}
\end{equation}

\begin{figure}[t!]
\begin{center}
\includegraphics[width=0.8\textwidth]{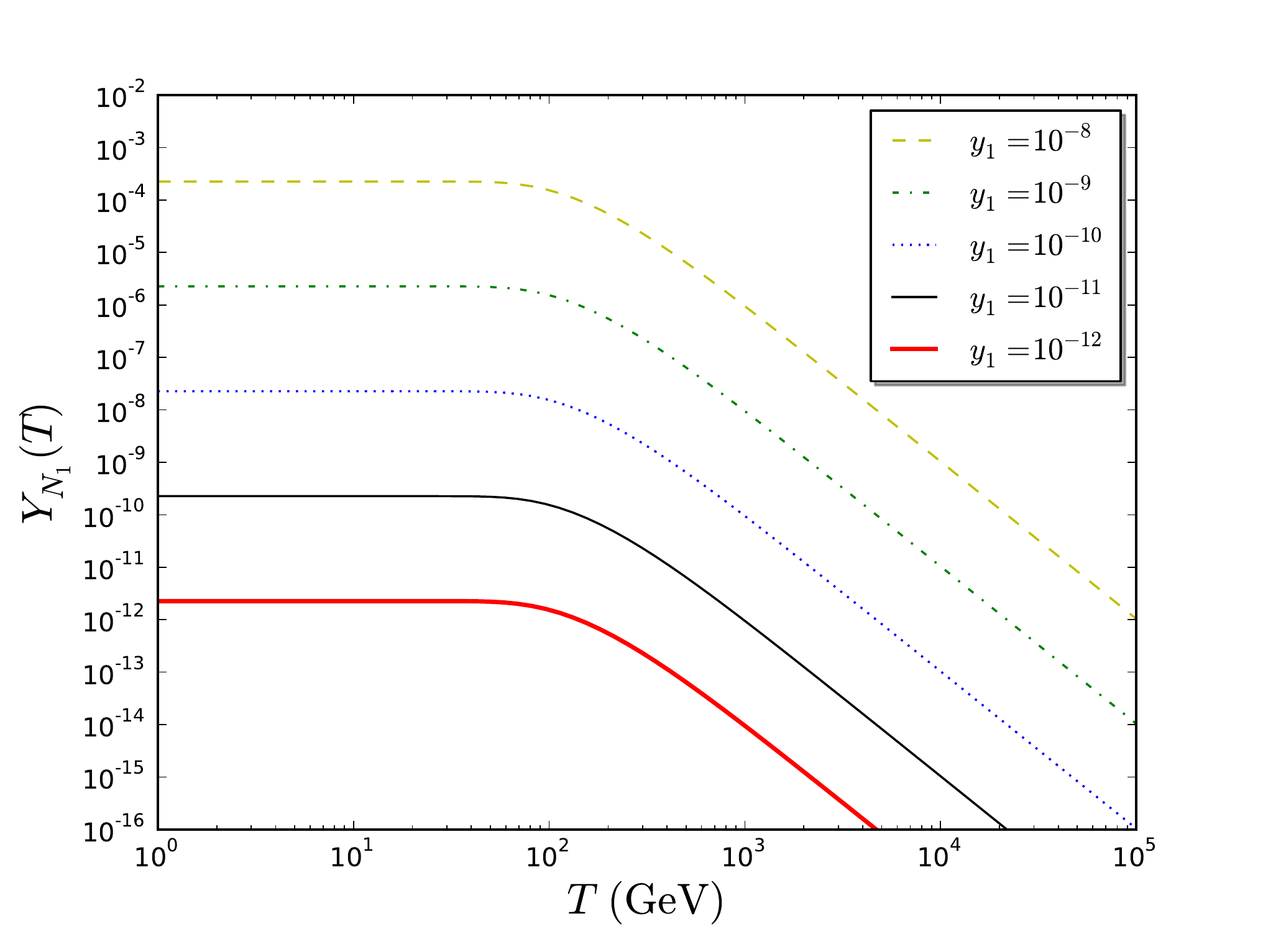} 
\caption{\small \it The dark matter yield due to the freeze-in process as a function of the temperature for different values of the FIMP Yukawa coupling $y_{1}$.  These results were obtained by solving the  Boltzmann equation, (\ref{BE}), for $m_S=400$ GeV. Notice that all other parameters ($M_i$, $y_{2,3}$) are irrelevant.}
\label{Fig1}
\end{center}
\end{figure}

We have studied quantitatively  the freeze-in production of dark matter in this scenario by solving numerically the Boltzmann equation (\ref{BE})
with the initial condition $Y_{N_1}=0$ for $T\gg m_S$. 
Figure~\ref{Fig1} shows the predicted dark matter abundance as a function of the temperature for different values of $y_1$.  The upper line corresponds to $y_1=10^{-8}$ and the lower one to $y_1=10^{-12}$. In this figure the common scalar mass, $m_S$, was set to $400~\mathrm{GeV}$. As stated before, the other parameters of the model are irrelevant. Notice, from the figure, that the abundance has the typical freeze-in behavior:  it increases steadily until the so-called freeze-in temperature is reached, remaining constant afterward.  Since the freeze-in temperature is determined by the mass of the decaying particle, it is the same for all the lines, as observed in the figure. Finally, the abundance is seen to depend  quadratically on $y_1$, as expected from 
equations~(\ref{GS1}), (\ref{GS2}) and (\ref{Yapprox}). 

\begin{figure}[t!]
\begin{center}
\includegraphics[width=0.8\textwidth]{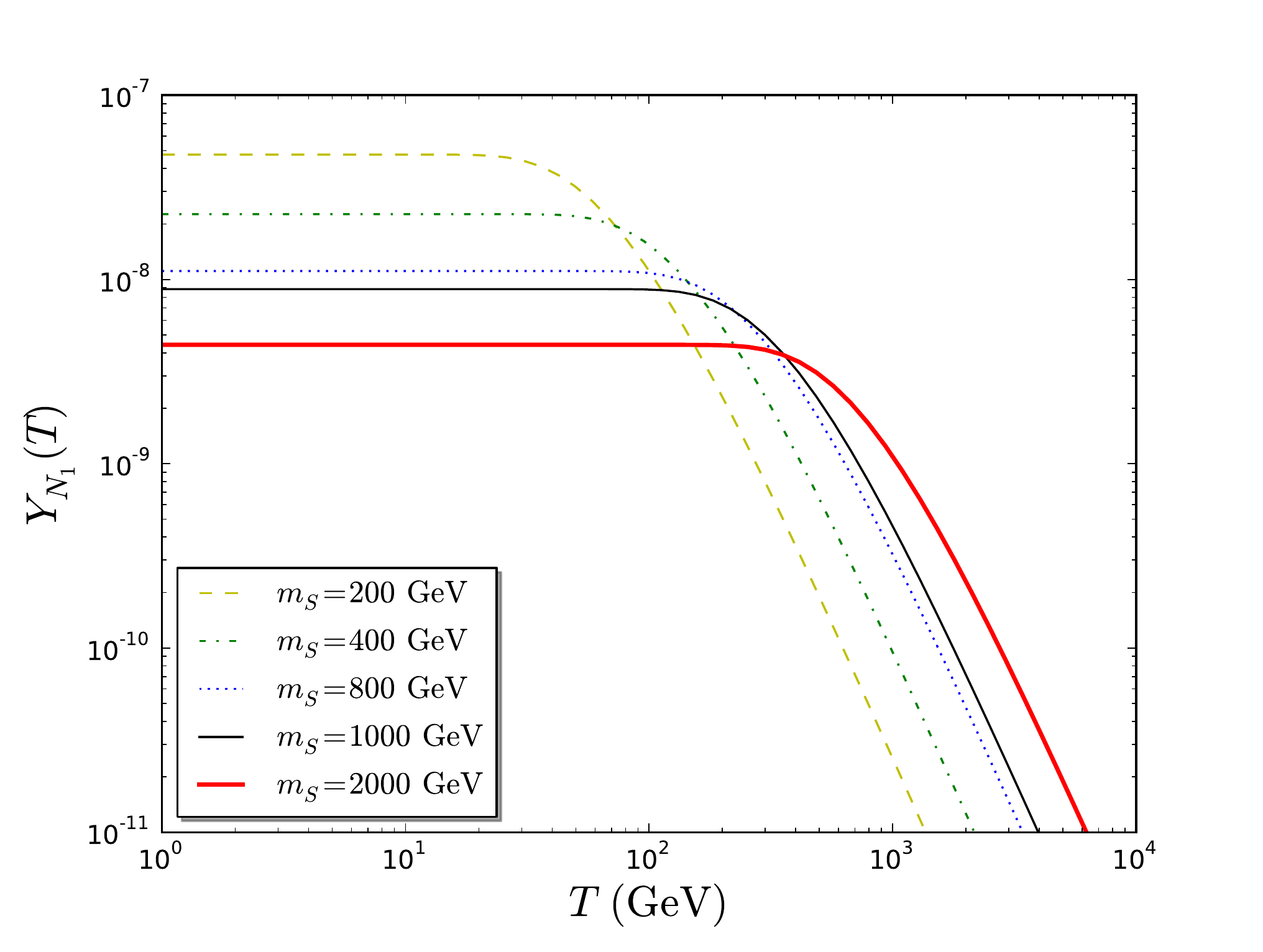} 
\caption{\small\it The dark matter yield due to the freeze-in process as a function of the temperature for different values of the common scalar mass, $m_S$. In this figure, the FIMP Yukawa coupling $y_{1}$ was set to $10^{-10}$. As before, all other parameters ($M_i$, $y_{2,3}$) are irrelevant. Notice that the freeze-in temperature depends on $m_S$.}
\label{Fig2}
\end{center}
\end{figure}

The dependence of $Y_{N_1}$ on $m_S$ is illustrated in figure~\ref{Fig2}, which displays the dark matter abundance as a function of the temperature for different values of $m_S$. 
In this figure $y_1$ was set equal to $10^{-10}$. One can clearly see that the freeze-in temperature increases with $m_S$, with the result  that the asymptotic value of $Y_{N_1}$ decreases with $m_S$. In fact, at low temperatures $Y_{N_1}$ is about ten times smaller for $m_S=2~\mathrm{TeV}$ than for $m_S=200~\mathrm{GeV}$.    
Notice from figures \ref{Fig1} and \ref{Fig2} that equation  (\ref{Yapprox}) is actually a very good approximation for the final yield obtained through the freeze-in mechanism.

\begin{figure}[t!]
\begin{center}
\includegraphics[width=0.8\textwidth]{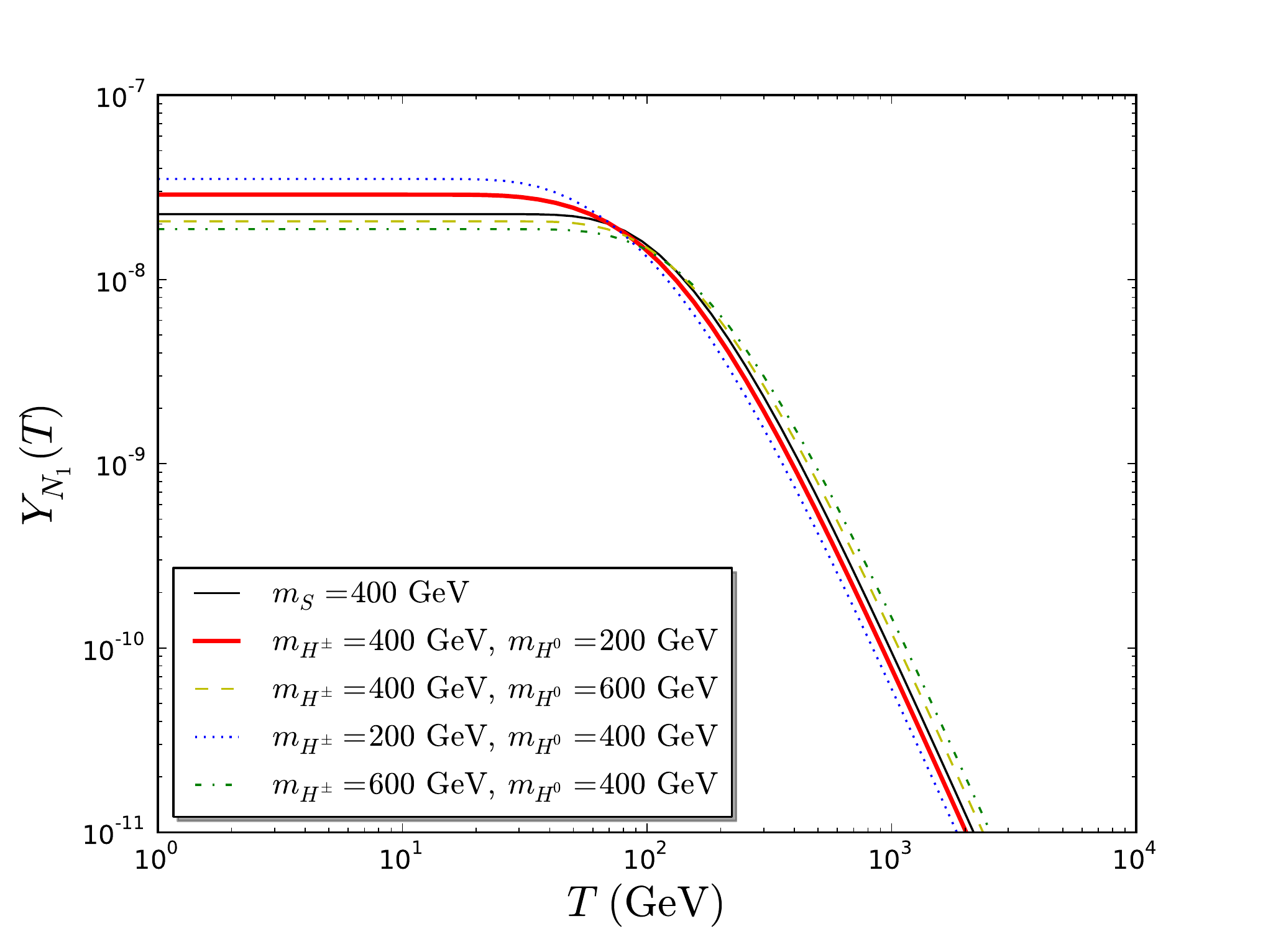} 
\caption{\small\it The dark matter yield due to the freeze-in process as a function of the temperature for different mass splittings among the odd  scalars. In this figure $y_{1}=10^{-10}$ and $m_{A^0}=400~\mathrm{GeV}$. As before, all other parameters ($M_i$, $y_{2,3}$) are irrelevant. }
\label{FigS}
\end{center}
\end{figure}

In the previous two figures we have assumed a common  mass, $m_S$, for all the odd scalars. In general, however, there will be a mass splitting between the three different states. To demonstrate that such mass splitting does not significantly affect our results, we show in figure~\ref{FigS} the dark matter abundance as a function of the temperature for different mass splittings. Notice that the variation in the final abundance due to the different kind of spectra is indeed very small. It is, therefore, a very good approximation to compute the dark matter abundance assuming that all odd scalars have the same mass $m_S$.  

The relic density of dark matter, $\Omega_{N_{1}}h^{2}$, is related to the asymptotic value of $Y_{N_{1}}$ at low temperatures by
\begin{equation}
	\Omega_{N_{1}}\,h^{2}\;=\; 2.744\times 10^{8}\, \frac{M_{1}}{\text{GeV}}\,Y_{N_{1}}(T_{0})\,,\label{OmegaN1}
\end{equation}
where $T_{0}=2.752$ K is the present day CMB temperature. It is this quantity that should be compared with the observed dark matter density as measured by WMAP \cite{Hinshaw:2012aka} and PLANCK \cite{Ade:2013lta}. For dark matter production via the freeze-in mechanism, the $N_1$ relic abundance  can be estimated as 
\begin{align}
  \Omega_{N_{1}}\,h^{2}&\approx 0.3\left(\frac{M_{1}}{0.1\,\mbox{GeV}}\right)\left(\frac{1\,\mbox{TeV}}{m_S}\right)\left(\frac{y_1}{10^{-10}}\right)^2,
\label{eq:rd}
\end{align}
where we used equations~(\ref{Yapprox}) and (\ref{OmegaN1}). Notice that this expression has the expected dependence on $m_S$, $y_1$ and $M_1$.

\begin{figure}[t!]
\begin{center}
\includegraphics[width=0.8\textwidth]{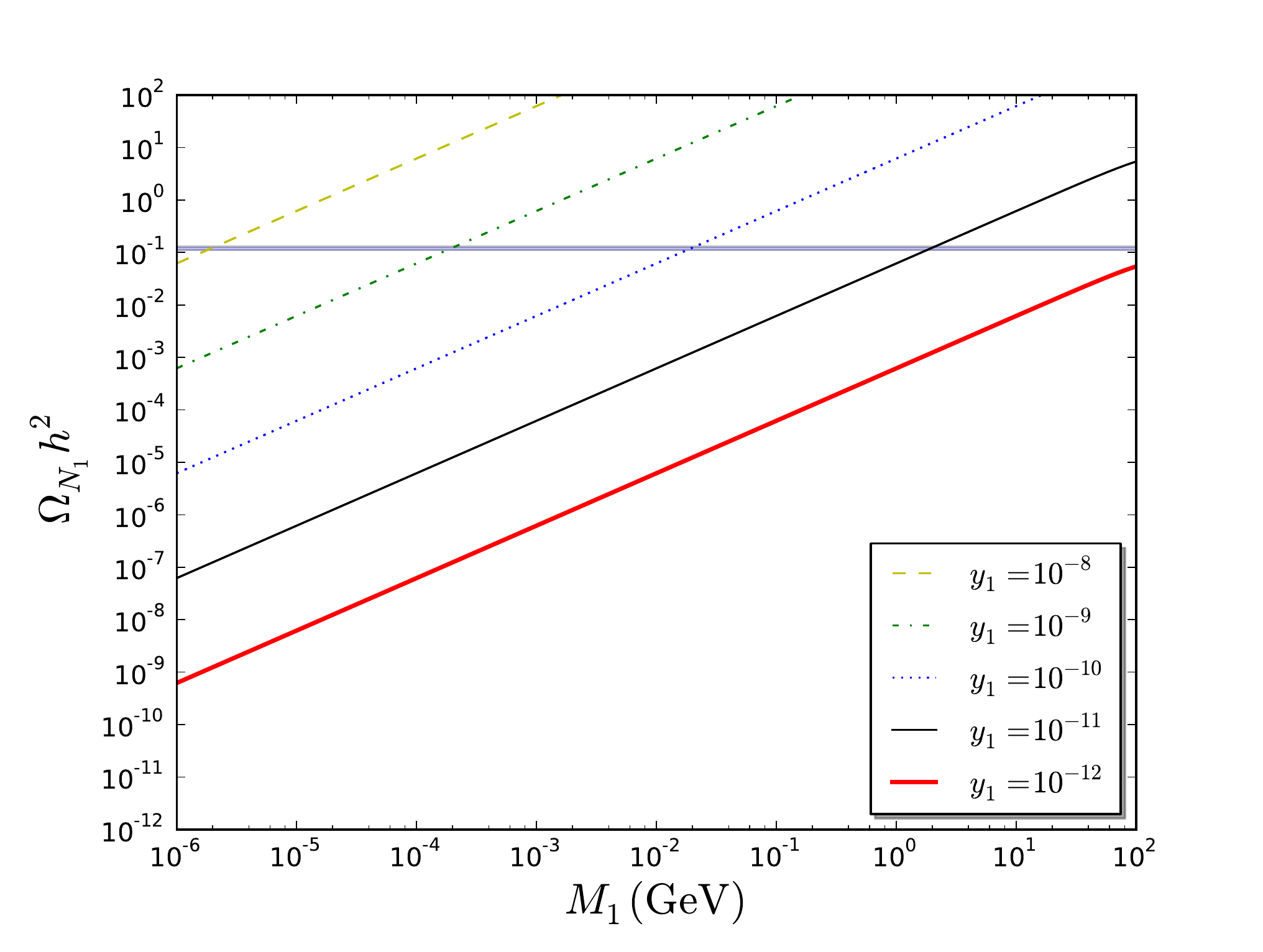} 
\caption{\small \it The freeze-in relic density as a function of the dark matter mass for different values of the FIMP Yukawa coupling $y_{1}$. We have fixed $m_S=400$ GeV in this figure.}
\label{Fig3}
\end{center}
\end{figure}

Figure \ref{Fig3} displays the $N_1$ relic density as a function of $M_1$ for $m_S=400~\mathrm{GeV}$ and different values of $y_1$. For $M_1$ we considered a minimum value of $1$ keV as indicated by  phase space density analysis \cite{Boyarsky:2008ju,Gorbunov:2008ka} and by the requirement of \emph{cold} or \emph{warm} dark matter. The maximum value was taken to be $100~\mathrm{GeV}$ in agreement with the idea that all odd particles live at or below the TeV scale. The horizontal band shows the region that is compatible with current observations. Notice that as we increase the mass a smaller value of $y_1$ is needed to be consistent with the data. Hence, whereas a keV particle requires $y_1\sim 10^{-8}$ a $100\GeV$ particle requires $y_1\sim 10^{-12}$.

\begin{figure}[t!]
\begin{center}
\includegraphics[width=0.8\textwidth]{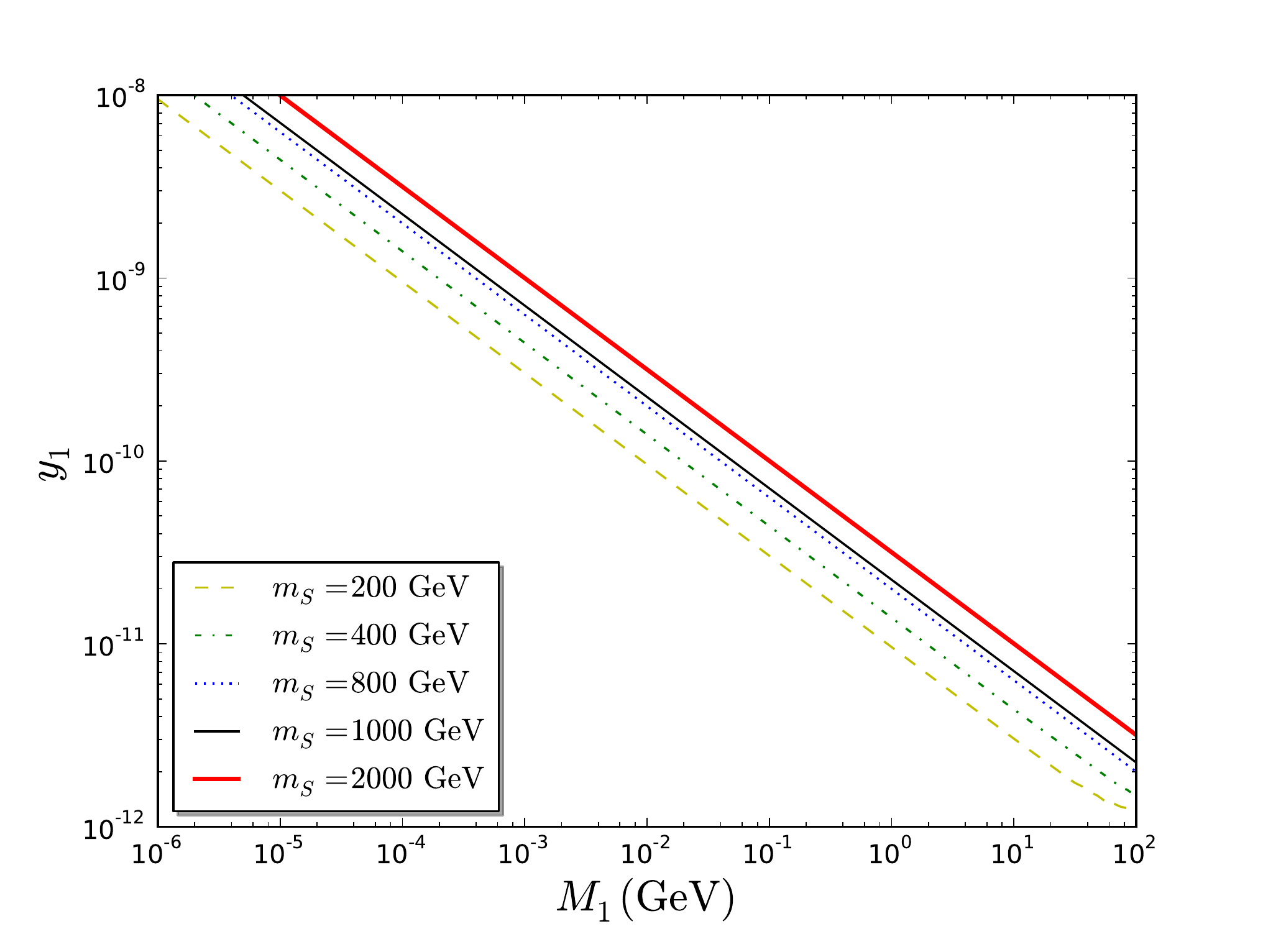} 
\caption{\small\it The regions in the plane ($M_1,y_1$) which give a freeze-in relic density in agreement with the observations. The lines correspond to different values of $m_S$.}
\label{Fig4}
\end{center}
\end{figure}

The viable parameter space for freeze-in dark matter in the scotogenic model is shown in figure~\ref{Fig4}. It displays, in the plane ($M_1$,$y_1$), the regions that are consistent with the observed dark matter density for different values of $m_S$. The freeze-in mechanism is thus able to explain the dark matter over a wide range of masses, from the keV to the TeV scale. Notice that at a given dark matter mass, the heavier $m_S$ the larger $y_1$. This figure is one of our main results, as it indicates the regions in the parameter space of the scotogenic model where the observed dark matter density can be accounted for entirely via freeze-in.  

If $N_1$ is very light, $M_1\sim 1-10~\mathrm{keV}$, the resulting dark matter is \emph{warm} rather than cold, with important implications for structure formation in the early Universe. Without freeze-in it is not possible to obtain warm dark matter in the scotogenic model because $N_1$ would thermalize and later decouple while relativistic, yielding a relic density about three orders of magnitude larger than observed. To make such scenario compatible with current observations would require either entropy dilution, e.g. via the decay of some other particle,  after $N_1$ production \cite{Ma:2012if} or a non-thermal production mechanism within a low reheating temperature scenario \cite{Ma:2012if}, both entailing significant departures from the model. Freeze-in provides instead a natural and simple way of obtaining warm dark matter in the scotogenic model.

If, in addition to freeze-in, other mechanisms  contribute to dark matter production, the lines in figure~\ref{Fig4} provide an upper bound on the coupling $y_1$ at a given value of $M_1$ and $m_S$.  As mentioned at the beginning of this section, in the scotogenic model the relic density of $N_1$ receives also a superWIMP contribution from the decays of the next-to-lightest odd particle after it has frozen out. Let us now turn our attention to that contribution.

\subsection{The superWIMP contribution}

In the superWIMP mechanism, the contribution to the dark matter relic density from the late decay of the next-to-lightest odd particle (NLOP from now on) is given by
\begin{equation}
	\Omega^{superWIMP}_{N_{1}}\,h^{2} \; = \; \frac{M_{1}}{M_{NLOP}}\, \Omega_{NLOP}^{freeze-out}\,h^{2}\,,
\label{swimp}
\end{equation} 
where $\Omega_{NLOP}^{freeze-out}\, h^{2}$ is the relic abundance, obtained via the usual freeze-out mechanism, of the NLOP . 
 In the scotogenic model, there are essentially two possibilities for the NLOP: $N_2$ or one of the scalars.  Next, we will in turn consider these two options.  

\mathversion{bold}
\subsubsection{$N_2$ as the NLOP}
\mathversion{normal}
If $N_2$ is the NLOP it will decay into dark matter via  the  scalar-mediated three-body process $N_{2} \to N_{1}\, \ell\, \overline{\ell}$. The requirement that this decay happens after the $N_2$ freeze-out (at $T\sim M_2/20$) implies that
\begin{equation}
	\Gamma(N_{2}\to N_{1}\, \ell \,\overline{\ell} ) \; \lesssim \; H(T\simeq M_{2}/20)\,. \label{cond1}
\end{equation}
This condition yields an upper bound on the product $y_1\,y_2$:
\begin{equation}
	y_{1}\,y_{2} \; \lesssim \;2\times 10^{-6} \left( \frac{m_{S}}{1\,\text{TeV}}\right)\left(\frac{1\,\text{TeV}}{M_{2}}\right)^{3/2}\,,
\end{equation}
which is always satisfied in this scenario  --the out-of-equilibrium condition gives a stronger bound.  
On the other hand, the lifetime of $N_{2}$ should be smaller than about $1$ second in order to not affect the Big Bang Nucleosynthesis (BBN) epoch. This requirement implies a lower bound on the product of the Yukawa couplings, namely
\begin{equation}
	y_{1}\, y_{2} \; \gtrsim \; 3 \times 10^{-12} \left( \frac{m_{S}}{1\,\text{TeV}} \right)^{2} \left(\frac{1\,\text{TeV}}{M_{2}}\right)^{5/2}\,.
\label{y1y2}
\end{equation}
At high values of the dark matter mass, this condition is very restrictive. If, for instance, $M_1\sim 100\GeV$ and $m_S,M_2\sim 1~\mathrm{TeV}$, it is not possible to satisfy  it as we know, from figure \ref{Fig4}, that $y_1$ should be no larger than about $10^{-12}$ (to avoid dark matter overproduction) and that $y_2$ cannot be of order $1$ due to the $\mu\to e\gamma$ bound. For $M_1\sim 1~\mathrm{keV}$ and the same values of $m_S$ and $M_2$, $y_1$ should be smaller than about $10^{-8}$ and the above bound is  satisfied for  $y_2\gtrsim 10^{-4}$. Taking $y_2\sim 10^{-2}$ as the upper limit on $y_2$ allowed by $\mu\to e\gamma$, the BBN constraint would exclude models with $y_1\lesssim 10^{-10}$  or equivalently with $M_1\gtrsim 100~\mathrm{MeV}$.  We can also use equation (\ref{y1y2}) to set a lower bound on the mass of $N_2$. Since $m_S\gtrsim 100~\GeV$, $y_1\lesssim 10^{-8}$ and $y_2\lesssim 10^{-1}$-$10^{-2}$, we get $M_2\gtrsim 10~\GeV$. Thus, the FIMP mechanism combined with the BBN constraint above tells us that $M_{2}$ and $M_{3}$ necessarily lie around the electroweak scale.

\begin{figure}[t!]
\begin{center}
\includegraphics[width=0.8\textwidth]{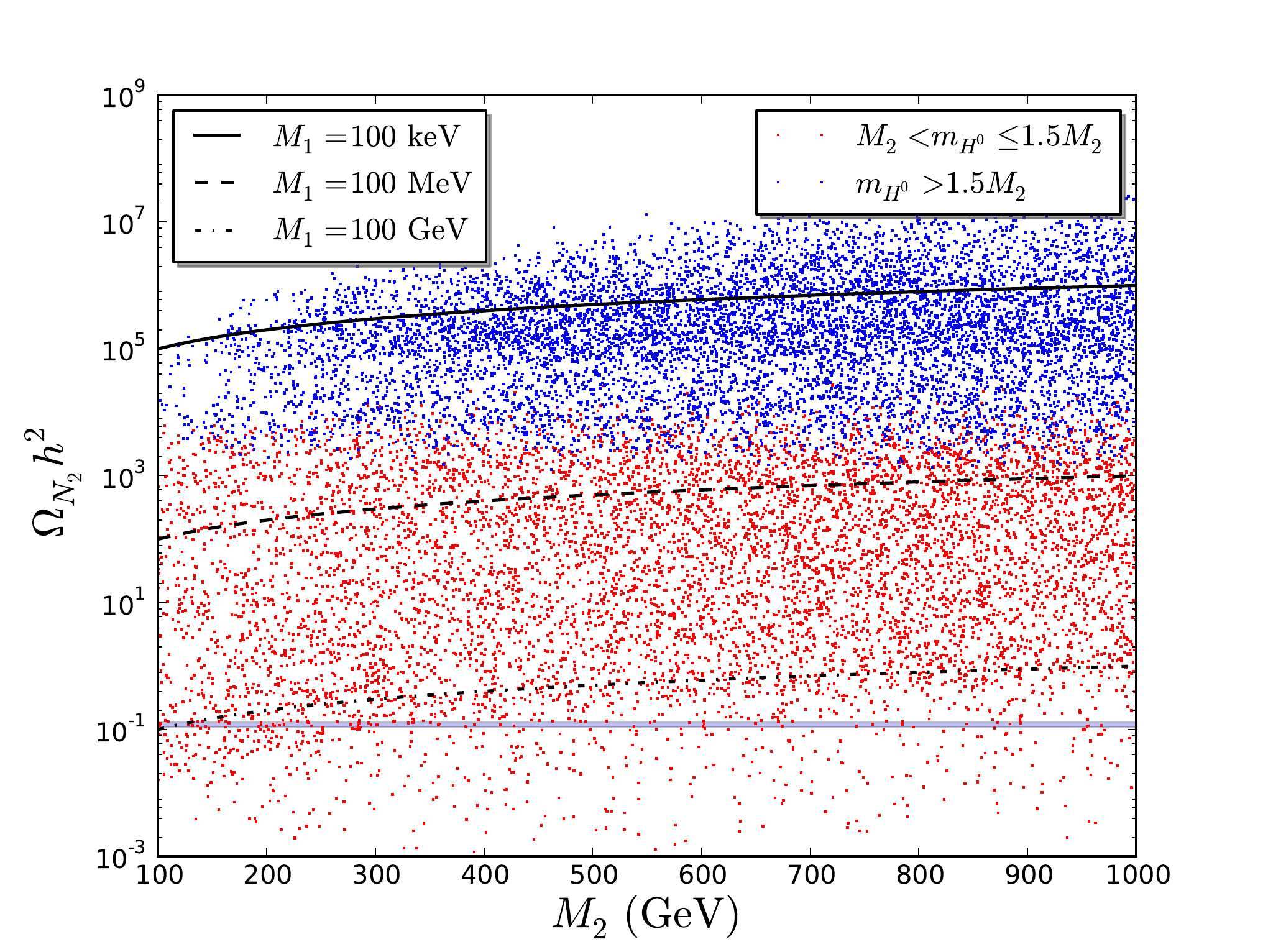} 
\caption{\small\it A scatter plot of the $N_2$ relic density, which is the result of a conventional freeze-out, versus $M_2$. For  this figure we have taken into account the bounds from neutrino masses, $\mu\to e\gamma$, and BBN. Notice that coannihilations with the scalars are relevant for the red  points but not for the blue ones.}
\label{FigscanN2}
\end{center}
\end{figure}

Regarding the value of $\Omega_{N_2}^{freeze-out}h^{2}$, previous studies have already shown that $N_2$-$N_2$ annihilations are not very efficient and usually require, to be consistent with the observed dark matter density, values of the Yukawas couplings so large that they run into conflict with the bounds from $\mu\to e\gamma$. Coannihilations between $N_2$ and  the scalars significantly help to increase the total annihilation rate, reducing  the relic density and  alleviating the tension with the $\mu\to e\gamma$ bound.  This situation is illustrated  in figure~\ref{FigscanN2}, which displays a scatter plot of the $N_2$ relic density versus $M_2$.  In it we have randomly varied all the parameters of the scotogenic model over a wide range: $1\,\mbox{keV}\leq M_1\leq 100\,\GeV$, $100\GeV\leq M_2\leq 1\,\mbox{TeV}$, $1\,\mbox{TeV}\leq M_3\leq 3\,\mbox{TeV}$, $M_2\leq m_{H_i }\leq 3\,\mbox{TeV}$, $10^{-12}\leq |Y^\nu_{\alpha 1}| \leq 10^{-8}$, $10^{-3}\leq \lambda_L\leq 1$.
All points in this figure  satisfy the constraints from neutrino masses, $\mu\to e\gamma$, and BBN. To precisely compute the relic density we used micrOMEGAs \cite{Belanger:2013oya}, which automatically includes all the relevant processes and takes care of possible resonant or coannihilation effects. With the goal of  isolating the effect of coannihilations, we have divided the sample into two sets according to the mass splitting between $N_2$ and the scalars. The mass-splitting is small for the red points (allowing for coannihilations) and large for  the blue points (excluding coannihilation effects). The horizontal band corresponds to the observed dark matter density. Notice that  coannihilations are essential to obtain a relic density in agreement with the observations. If $m_{H^0}>1.5~M_2$ (blue points), the $N_2$ relic density after freeze-out is always very large --at least four orders of magnitude larger than the observed dark matter density. Thus, compatibility  with current data requires $M_1/M_2\lesssim 10^{-4}$, according to equation~(\ref{swimp}). And since $M_2$ is at most of order TeV,  $M_1$ necessarily lies below the GeV scale. A large hierarchy between $M_1$ and $M_2$ is thus an essential  condition in this scenario.   If, on the contrary,  $M_{1}<m_{H^0}\leq1.5~M_2$ (red points), the $N_2$ relic density can even reach values below the observations. Consequently, no strict bounds on $M_1/M_2$ can be derived based on the relic density.     

If the dark matter density were dominated by the superWIMP contribution, $\Omega^{superWIMP}_{N_1}\,h^{2}\gg \Omega^{freeze-in}_{N_1}\,h^{2}$, one would need to ensure that the dark matter is non-relativistic at the onset of structure formation; otherwise it would behave as hot dark matter. This condition leads to a relation between the  $N_2$ decay time and the ratio $M_1/M_2$. A detailed analysis of this issue  can be found in \cite{Gelmini:2009xd}.  They found, in particular, that one can obtain \emph{warm} dark matter for $24~\mathrm{keV}\lesssim M_1\lesssim 24~\mathrm{MeV}(M_2/100~\mathrm{GeV})$. In figure~\ref{FigscanN2} we have also displayed, for three different values of $M_1$ (100 GeV, 100 MeV, 100 keV), the regions where the superWIMP contribution accounts for the entire dark matter density. If $M_1=100~\mathrm{MeV}$, for example,  then along the dashed line the superWIMP contribution agrees with the observed relic density. Models above that line are excluded (for that value of $M_1$) as they overproduce dark matter whereas models below that line require the freeze-in contribution to be compatible with the data. Even though the relic density constraint is satisfied along the dashed-dotted line for $M_1=100\GeV$, that value of $M_1$ is actually ruled out by the BBN bound, as explained before. If  $M_1=100~\mathrm{keV}$, the superWIMP contribution accounts for the dark matter along the solid line and one obtains warm dark matter.

\subsubsection{A scalar as the NLOP}
If one of the scalars is the NLOP, its direct decay  into $N_{1}$ and SM leptons after  decoupling from the thermal bath will give an additional contribution to the dark matter abundance. The condition that the decay takes place after the scalar freeze-out but before BBN translates into
\begin{equation}
	 10^{-13}\left(\frac{1\,\text{TeV}}{m_{S}} \right)^{1/2}\;\lesssim\; y_{1} \;\lesssim \; 10^{-8}\left(\frac{m_{S}}{1\,\text{TeV}}\right)^{1/2}\,,
\end{equation}
where we have implicitly assumed that the decaying scalar and $N_1$ are not highly degenerate. These bounds are easily satisfied for the range of parameters relevant for freeze-in --see figure~\ref{Fig4}. For concreteness, in the following we assume the NLOP scalar to be $H^0$ but it must be kept in mind that the results for the other scalars are similar.   

\begin{figure}[t!]
\begin{center}
\includegraphics[width=0.8\textwidth]{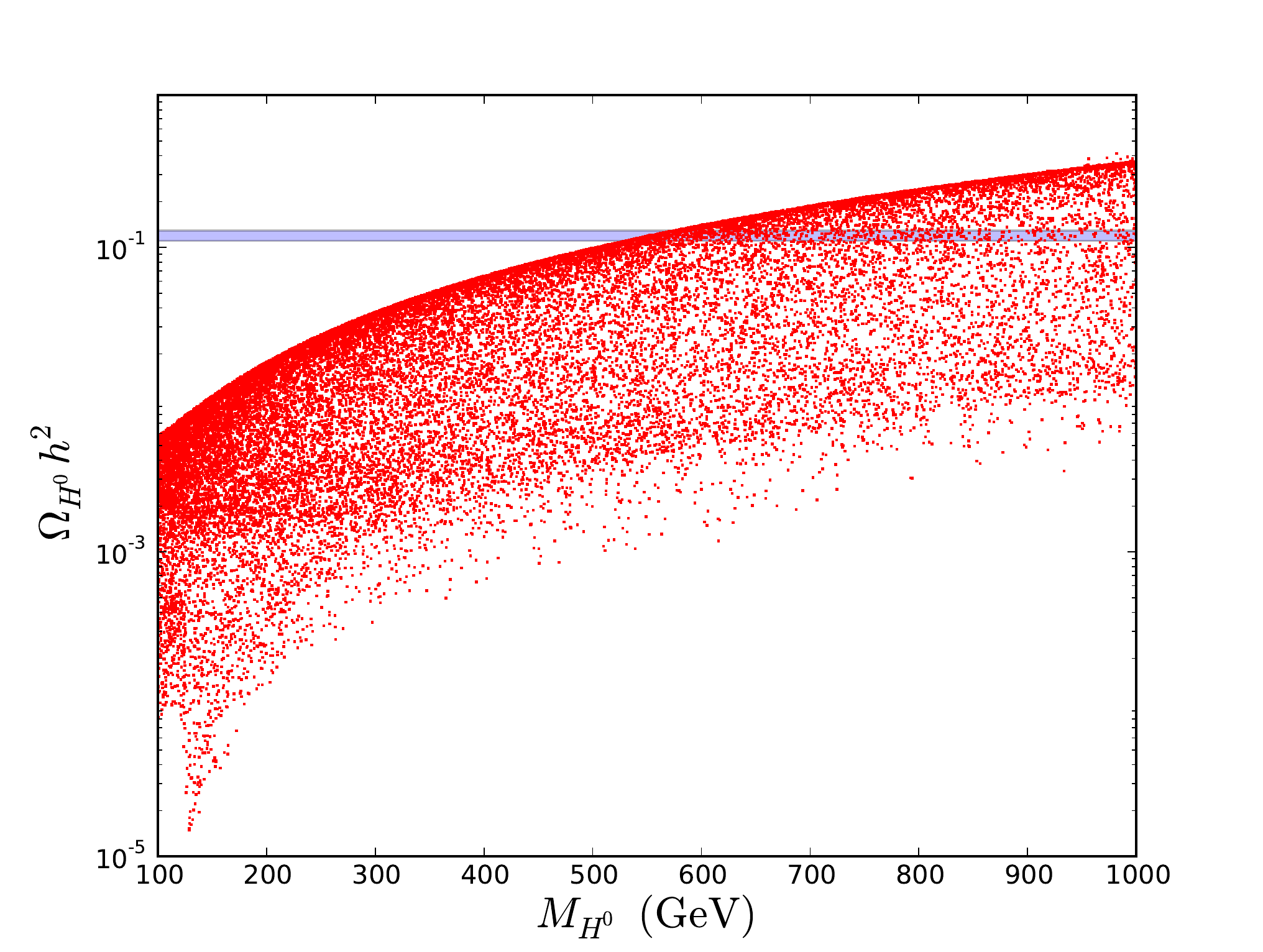} 
\caption{\small\it A scatter plot of the $H^0$ relic density, which is the result of a conventional freeze-out, versus $m_{H^0}$. For  this figure we have taken into account the bounds from neutrino masses, $\mu\to e\gamma$, and BBN.}
\label{FigscanH0}
\end{center}
\end{figure}

The relic density of $H^0$ in this scenario is very much alike that in the inert-doublet model.  A remarkable feature of this model is that if $M_W<m_{H^0}\lesssim 500\GeV$ the relic density is always too small to satisfy the dark matter constraint. The reason being that the annihilation into gauge bosons are so efficient that they deplete the dark matter abundance well below the observed value. Only for masses above $500$ GeV (or  below $M_W$) it is possible to satisfy the dark matter bound. Figure \ref{FigscanH0} shows a scatter plot of the $H^0$ relic density versus $m_{H^0}$ obtained after  varying all the parameters of the scotogenic model ($m_{H_i}\leq M_2\leq 1\,\mbox{TeV}$, $M_2\leq M_3\leq 3\,\mbox{TeV}$, $100\GeV\leq m_{H_i }\leq 1\,\mbox{TeV}$  and the others as before) and selecting those consistent with neutrino masses, $\mu\to e\gamma$, and BBN. As before, the horizontal band shows the observed dark matter density. Notice from the figure that the relic density increases with the mass and that, as expected, it only crosses the experimental value for masses above $500$ GeV or so. This fact has a very important implication: if $m_{H^0}<500\GeV$ the superWIMP contribution to the relic density is negligible and the entire dark matter density has to be explained via the freeze-in mechanism. That is, in contrast to the case where $N_2$ is the NLOP, we can identify an important region of the parameter space, $m_{H^0}<500\GeV$, where the freeze-in contribution is always dominant.

If $m_{H^0}>500\GeV$, the superWIMP contribution could be the dominant one.  In that case, since the $H^0$ relic density is never much larger than the observed dark matter density, a mild hierarchy between $N_1$ and $H^0$ is required, $m_{H^0}/M_{N_1}\lesssim 4$ (see figure~\ref{FigscanH0}). Even for such large values of $m_{H^0}$, however, the freeze-in contribution can dominate the $N_1$ relic density.  

\subsection{Implications}
As we have seen, FIMP dark matter can indeed be realized in the scotogenic model. In general, the relic density of $N_1$ is the sum of a  freeze-in contribution and a contribution  from the decay of the NLOP. Whether one or the other dominates depends strongly on the parameters of the model.  Let us now briefly discuss the implications of this scenario.

A generic prediction of FIMP models is that the NLOP, which must decay into the FIMP, is very long-lived \cite{Hall:2009bx}, providing a possible way of testing these scenarios at colliders such as the LHC.
In the scotogenic model, the most interesting  signal occurs when the charged scalar is the NLOP. In that case the relic density is expected to be dominated by the freeze-in process and, from  equation (\ref{eq:rd}), we have that
\begin{equation}
y_1^2  = 4\times10^{-20} \left(\frac{0.1~\mathrm{GeV}}{M_1}\right) \left(\frac{m_{H^+}}{1~\mathrm{TeV}}\right).
\end{equation}
Now, let us suppose that this charged scalar, with a mass in the range [100 GeV, 1 TeV], is produced at the LHC.
Its decay width is given by equation (\ref{GS2}). Therefore, taking into account the value of $y_1$ derived above we obtain
\begin{equation}
\Gamma(H^+\to \ell^+ N_1) = \frac{1}{4\pi}\, 10^{-17}~\mathrm{GeV} \left(\frac{0.1~\textrm{GeV}}{M_1}\right) \left(\frac{m_{H^+}}{1~\mathrm{TeV}}\right)^2 .
\end{equation}
Thus, the $H^+$ decay length, $l(H^+)$, is (ignoring for the moment the Lorentz boost factor)
\begin{align}
 l(H^+) &=  3\times10^5 \mathrm{cm} \left(\frac{M_1}{1~\mathrm{GeV}}\right) \left(\frac{1~\mathrm{TeV}}{m_{H^+}}\right)^2, \\
&\lesssim 3~\mathrm{meters} \left(\frac{1~\mathrm{TeV}}{m_{H^+}}\right)^2\quad\mbox{for}\quad M_1\lesssim 1~\mathrm{MeV}.
\end{align}
Including the Lorentz boost factor amounts to multiplying this upper limit by a factor from $2$ to $7$. Thus, for dark matter masses in the range [10 keV, 1 MeV]  the decay length is below 10 meters and $H^+$ decays  inside the detector, leaving a charged lepton plus missing energy signature that could be searched for at the LHC. If the decay happens instead outside the detector, evidence for $H^+$ could be found at the LHC via searches for long-lived charged particles \cite{Chatrchyan:2013oca}. It is beyond the scope of the present paper, however, to determine whether these signals can actually be used to set meaningful constraints on this scenario.

Another generic feature of FIMP dark matter is the absence of signals at direct or indirect detection experiments --a direct consequence of the feeble interactions that are required to prevent the dark matter from reaching thermal equilibrium in the early Universe. These experiments provide, nonetheless, an unambigous way of falsifying this scenario: as soon as a positive signal is confirmed in any dark matter detection experiment we would learn  that dark matter does not consist of FIMPs and more specifically that the scenario we studied in this section is ruled out. Such signal would instead give a strong support to the WIMP paradigm of dark matter. But if the next generation of dark matter experiments, such as XENON1T \cite{Aprile:2012zx}, fails to find evidence of dark matter, the WIMP framework would be in trouble and alternative scenarios that can naturally explain the absence of such evidence would become much more appealing. In that hypothetical  future the  FIMP scenario could become the standard framework to account for the dark matter. Only time will tell which of these two possible outcomes regarding dark matter detection will actually be realized.

\section{FIMP decay into dark matter}
\label{sec:decay}
In the previous section we assumed that the singlet fermion that does not reach thermal equilibrium in the early Universe ($N_1$) was also the lightest particle odd under the $Z_2$ symmetry,  and consequently the dark matter candidate. It may well be though that $N_1$ is not the lightest odd particle so that the dark matter candidate is instead one of the neutral scalars or another singlet fermion.  In that case, $N_1$ is unstable and decays into the dark matter, increasing its relic density. Thus, $N_1$ modifies the regions where the dark matter constraint is satisfied, allowing for regions which in the usual freeze-out scenario are under-dense ($\Omega^{freeze-out}h^{2}<0.1$) to become compatible with the observed dark matter density. Since the singlet (say $N_2$) relic density obtained via freeze-out is typically larger than the observed one, see e.g. figure~\ref{FigscanN2}, an additional contribution from FIMP decay is usually not welcome as it will only help in very specific cases.  Much more  interesting is the situation where one of the neutral scalars is the dark matter candidate,  for we know that over a significant region of the parameter space its freeze-out relic density is very small --see e.g. figure~\ref{FigscanH0}. For definiteness, we take $H^0$ as the dark matter particle and assume that all the odd scalars are lighter than $N_1$,  $m_{H^0}<m_{A^0},m_{H^\pm}<M_1$.  Notice that, contrary to the discussion in the previous section, the dark matter particle in this case is a WIMP.

The $H^0$ relic density will receive two contributions, one from freeze-out and one from the late decay of $N_1$. We can then write
\beq
\Omega_{H^0}\,h^{2}=\Omega_{H^0}^{freeze-out}\,h^{2}+\Omega_{H^0}^{N_1-decay}\,h^{2}
\eeq
with 
\beq
\Omega_{H^0}^{N_1-decay}\,h^{2}=\frac{m_{H^0}}{M_1}\,\Omega_{N_1}^{freeze-in}\,h^{2}.
\eeq
Let us now proceed to calculate $\Omega_{N_1}^{freeze-in}h^{2}$ in this case. The dominant freeze-in production process is the inverse decay of $N_1$, $X+\ell\to N_1$, where $X$ denotes an odd scalar and $\ell$ is a SM lepton.  The $N_1$ yield, $Y_{N_{1}}(T)=n_{N_1}(T)/s(T)$, is computed by solving the same Boltzmann equation as in the previous section, equation~(\ref{BE}), but with a different production rate
\begin{equation}
	\gamma_{N_{1}}(T)\;=\;	\sum\limits_{X}\frac{g_{N_1}\,M_{1}^{2}\, T}{2\,\pi^{2}}\,K_{1}\left(M_1/T\right)\,\Gamma\left(N_1\to X \, \ell\right),
\end{equation}    
where $g_{N_{1}}=2$ because $N_1$ is a Majorana fermion.
The decay width for the three decay channels of $N_1$ into scalars are given by 
\begin{align}
\Gamma(N_1\to H^0/A^0\,\nu_\alpha)&=\frac{(M_1^2-m_{H^0/A^0}^2)^2}{64\pi M_1^3}\left| Y^{\nu}_{\alpha 1} \right|^{2},\\
\Gamma(N_1\to H^+\,\ell_\alpha)&=\frac{(M_1^2-m_{H^+}^2)^2}{32\pi M_1^3}\left| Y^{\nu}_{\alpha 1} \right|^{2}.
\end{align}
Hence, the total decay rate of $N_{1}$ is
\begin{align}\label{eq:GammaN1}
\Gamma_{N_1}&=\frac{M_1}{8\pi}(1-m_S^2/M_1^2)^2\left(\sum\limits_{\alpha} \left|Y^{\nu}_{\alpha 1} \right|^{2}\right)\approx \frac{M_1}{8\pi}\sum\limits_{\alpha} \left|Y^{\nu}_{\alpha 1} \right|^{2}\,,
\end{align}
where the last approximation is valid unless $N_1$ is highly degenerate with the scalars.

The abundance $Y_{N_1}$ at certain temperature $T$ can then be expressed as
\begin{align}
Y_{N_1}(T)&=8.49\times10^{17}\mbox{GeV}\,M_1^2\,g_1\,\Gamma_{N_1}\int_{T}^{T_i}\frac{K_1(M_1/T)}{g_s(T)\sqrt{g_{\rho}(T)}T^5}dT.
\end{align}
whereas the $N_1$ relic density  is 
\begin{align}
\Omega_{N_1}^{freeze-in}h^2&=2.33\times10^{26}\,M_1^3\,g_1\,\Gamma_{N_1}\int_{T_0}^{T_i}\frac{K_1(M_1/T)}{g_s(T)\sqrt{g_{\rho}(T)}T^5}dT.
\end{align}
Finally, we can approximate $\Omega_{H^0}^{N_1-decay}$ as
\begin{align}
\Omega_{H^0}^{N_1-decay}h^2&\approx0.1\left(\frac{m_{S}}{100\,\mbox{GeV}}\right)\left(\frac{1\,\mbox{TeV}}{M_1}\right)\left(\frac{y_1}{2\times10^{-12}}\right)^2.
\end{align}
Thus, a coupling of order $10^{-12}$ is required to  account for the entire dark matter density via the decay of $N_1$. 

The above result holds provided that $N_1$ decays  after the $H^0$ freeze-out, $\Gamma_{N_1}\lesssim H(T_{H^0}^{f.o.})$. Since the  $N_1$ decay rate is given by 
\begin{align}\label{eq:BBN}
\Gamma_{N_1}&\sim 4.8\times10^{-22}\,\mbox{GeV}\left(\frac{M_1}{1\,\mbox{TeV}}\right)\left(\frac{y_1}{2\times10^{-12}}\right)^2
\end{align}
whereas 
\begin{align}\label{eq:foH0}
H(T^{f.o.}_{H^0})= H(m_S/x_{f.o.})\sim 3.4\times 10^{-17}\,\mbox{GeV}\left(\frac{m_S}{100\,\mbox{GeV}}\right)^2\left(\frac{20}{x_{f.o.}}\right)^2.
\end{align}
one can see that this condition is easily satisfied. In order to not alter the predictions of BBN, one must also ensure that  $\Gamma_{N_1}\gtrsim 1/0.3$ sec$^{-1}=2.2\times 10^{-24}$ GeV, which  is seen to be fulfilled for the values required to obtain the correct dark matter density. 

\begin{figure}[t!]
\begin{center}
\includegraphics[width=0.8\textwidth]{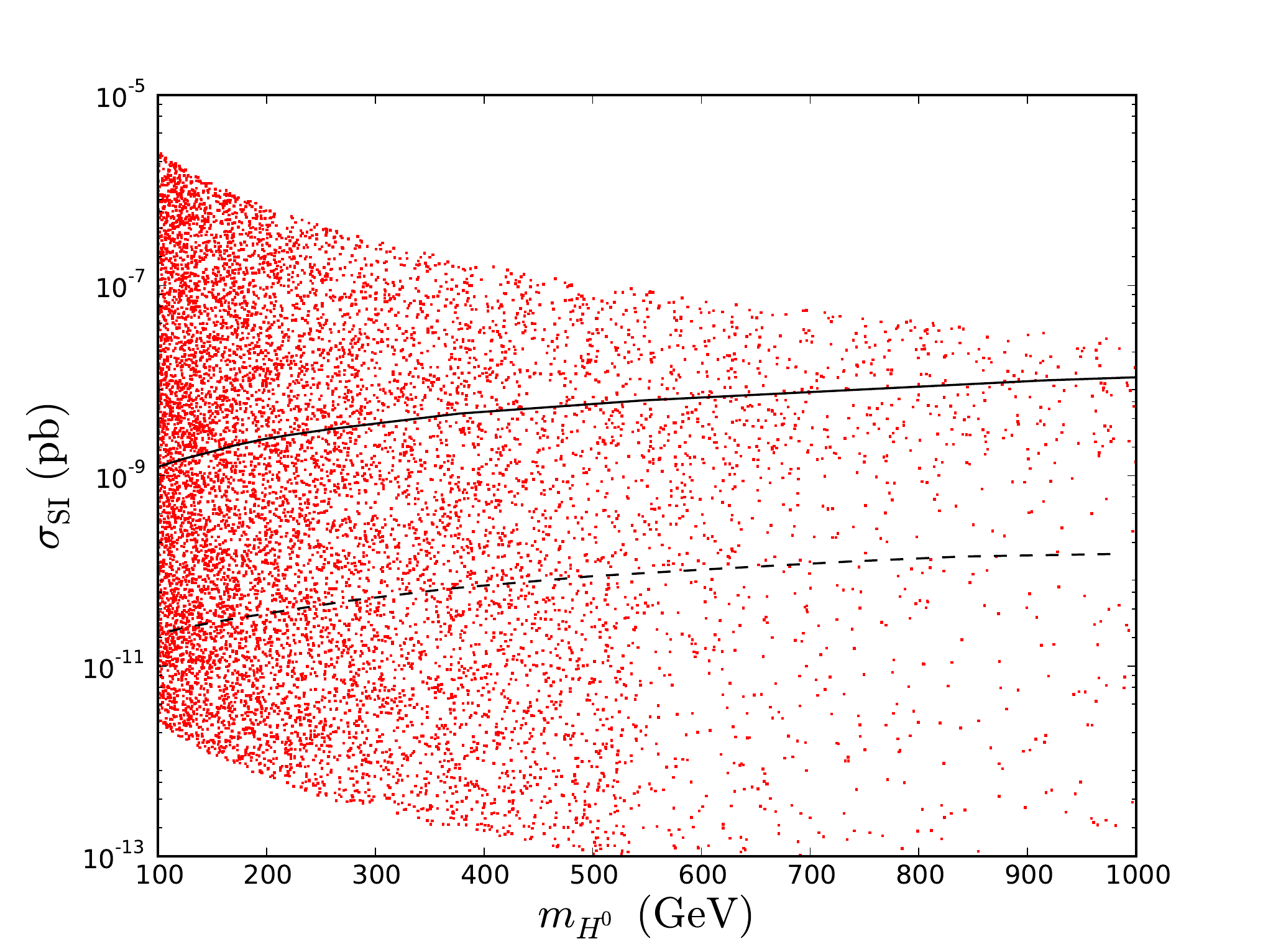} 
\caption{\small\it A scatter plot of the spin-independent direct detection cross section  versus $m_{H^0}$. The solid line shows the current bound by the LUX experiment \cite{Akerib:2013tjd} whereas the dashed line displays the expected sensitivity of XENON1T \cite{Aprile:2012zx}.}
\label{FigscanSI}
\end{center}
\end{figure} 

The idea then is that if $\Omega_{H^0}^{freeze-out}h^{2}<\Omega_{\text{DM}}h^{2}$ we can always choose a  value of $y_1$ such that the contribution from the decay of $N_1$ compensates for the deficit and one gets  a relic density in agreement with the observations, $\Omega_{H^0}=\Omega_{\text{DM}}$. That is, the presence of the FIMP allows us to enlarge the viable parameter space of the model by rescuing those regions  where freeze-out gives a too small relic density. In particular, the region $m_{H^0}\lesssim 500\GeV$ becomes viable within this setup. 

\begin{figure}[t!]
\begin{center}
\includegraphics[width=0.8\textwidth]{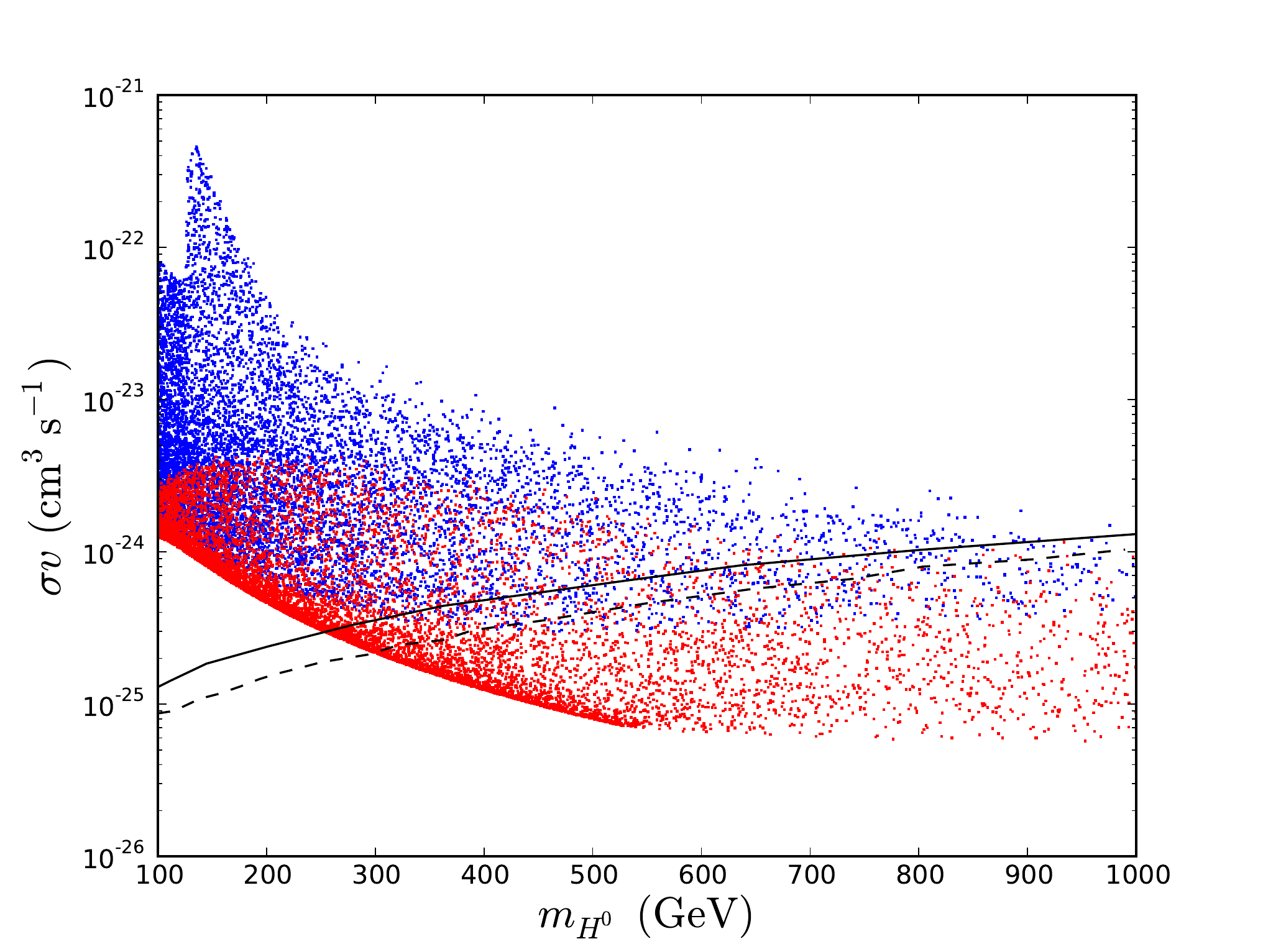} 
\caption{\small\it A scatter plot of the dark matter annihilation rate today ($\sigma v$)  versus $m_{H^0}$. The two lines show current bounds obtained by the Fermi-LAT collaboration for annihilation into $b\bar{b}$ (solid line) and $W^+W^-$ (dashed line).}
\label{Figscansv}
\end{center}
\end{figure}

 The resulting scenario is quite similar to that discussed in \cite{Klasen:2013jpa}. The difference being the mechanism that allows to increase the relic density. In \cite{Klasen:2013jpa} it was the coannihilations with the singlet fermions whereas in our case is the late decay of the FIMP.

Since the dark matter particle $H^0$ is a WIMP, the usual direct and indirect detection signals are expected and one must make sure that current bounds are respected. The dark matter phenomenology of $H^0$ is reminiscent of that  in the inert doublet model. Direct detection, for instance, proceeds via a Higgs mediated diagram and is determined by  the  coupling $\lambda_L=(\lambda_{3}+\lambda_{4}+\lambda_{5})/2$. Figure \ref{FigscanSI} shows a scatter plot of the spin-independent direct detection cross section versus the dark matter mass. The figure was obtained after randomly varying the different parameters of the model ($1\,\mbox{TeV}\leq M_1\leq 3\,\mbox{TeV}$, $M_1\leq M_2\leq 3\,\mbox{TeV}$, $M_2\leq M_3\leq 3\,\mbox{TeV}$, $100\GeV\leq m_{H_i }\leq 1\,\mbox{TeV}$ and the others as before) and imposing the known experimental bounds (neutrino masses, $\mu\to e\gamma$, etc.).  The $H^0$ relic density is consistent with the observed dark matter density thanks to the contribution from $N_1$ decays. For comparison we show the current experimental bound (solid line) \cite{Akerib:2013tjd} and the expected sensitivity of future experiments (dashed line) \cite{Aprile:2012zx}. Even though several models are already excluded (those above the solid line) and many more will be probed by future experiments (those above the dashed line), one can still find models with small values of $\sigma_{\rm SI}$ over the entire range of masses we explore. Direct detection bounds therefore do not restrict the range of the dark matter mass in this scenario.

Unsurprisingly, the indirect detection bounds turn out to be more constraining. Indeed, since $\Omega_{H^0}^{freeze-out}h^{2}\ll\Omega_{\rm DM}h^{2}$, we expect  annihilation rates larger than those typically associated with  WIMPs, $\sigmav_{H^0}\gg\sigmav_{thermal}\sim 3\times 10^{-26}~\text{cm$^{-3}$/s}$. Figure \ref{Figscansv} shows a scatter plot of $\sigma v$ versus the dark matter mass. As before, the correct relic density is obtained via $N_1$ decays and the experimental bounds were taken into account. In blue we show the models that are excluded by the direct detection bound on $\sigma_{\rm SI}$ --see figure \ref{Figscansv}.  In red  we show instead the points that are  consistent with that bound.    
Notice that $\sigma v$ can indeed be much larger than the so-called thermal value.   The solid and dashed lines show the current bounds obtained by the   Fermi-LAT collaboration, 
for dark matter annihilation into $b$ quarks \cite{Ackermann:2013yva} and $W$ gauge bosons \cite{Ackermann:2011wa}, respectively. They exclude all models with $m_{H^0}\lesssim 300\GeV$. For higher values of the dark matter mass, $m_{H^0}\gtrsim 300\GeV$, one can easily find models compatible with both direct and indirect detection constraints.

In contrast to the scenario with FIMP dark matter discussed in the previous section, this setup, where $H^0$ is the dark matter particle and the decay of $N_1$ contributes to its relic density, can  be probed by both  direct and indirect detection experiments. And as we have seen, the expected signals are generally significant. 

\section{Conclusions}
\label{sec:con}

We have shown that in  the scotogenic model --one of the simplest extensions of the SM that can account for neutrino masses and dark matter at the TeV scale-- one (and only one) of the singlet fermions, $N_1$, can be out of equilibrium in the early Universe and behave as a FIMP, with important implications for the phenomenology of this model. This setup predicts, for instance, that one of the light neutrinos is essentially massless. Within this framework the dark matter candidate can be a FIMP, $N_1$,  or a WIMP, $H^0$. In the former case, the relic density of dark matter receives two contributions, one from freeze-in and another one from the late decay of the next-to-lightest odd particle --the superWIMP contribution. The freeze-in contribution was found to be dominated by the decays of the scalars and its dependence with the different parameters of the model was examined in detail. Specifically, we determined the regions in the plane ($M_1$, $y_1$) where freeze-in can account for the observed dark matter density and found that they span a wide range of masses, from the keV to the TeV scale. The superWIMP contribution was also discussed and shown to strongly depend on the identity of  the next-to-lightest odd particle. In the latter case, when $H^0$ is the dark matter particle, the relic density is not only the result of a freeze-out but receives and additional contribution  from the late decays of $N_1$. This second contribution allows to increase the dark matter relic density, opening up new viable regions in the parameter space of the model. Thanks to this contribution from $N_1$ decay, regions that within the standard scenario feature a too small relic density, such as $m_{H^0}\lesssim 500\GeV$, can become compatible with the observed dark matter density. We demonstrated that in this case one generally expects observable signals at direct and indirect dark matter experiments.      

\section*{Acknowledgments}
The work of E. M. is supported by the ERC Advanced Grant project ``FLAVOUR'' (267104). E. M. is grateful to  Nordita -- Nordic Institute for Theoretical Physics for the kind hospitality during the final phase of this work.
C.Y. is partially supported by the ``Helmholtz Alliance for Astroparticle Physics HAP''
 funded by the Initiative and Networking Fund of the Helmholtz Association. O.Z. has been partially supported by Sostenibilidad-UdeA, UdeA/CODI grant IN624CE and COLCIENCIAS through the grant number 111-556-934918. O.Z. also acknowledges support from the German Academic Exchange Service (DAAD).

\bibliographystyle{hunsrt}
\bibliography{darkmatter}

\end{document}